\documentclass[journal=jacsat,manuscript=article]{achemso}

\usepackage[version=3]{mhchem} % Formula subscripts using \ce{}
\usepackage[separate-uncertainty=true, per-mode=symbol]{siunitx}
\usepackage{soul}
\usepackage{amsmath}
\author{Shima Rajabali}
\email{shimar@phys.ethz.ch}
\author{Josefine Enkner}
\affiliation[ETH]
{Institute of Quantum Electronics, ETH Z{\"u}rich, 8093 Z{\"u}rich, Switzerland}
\author{Erika Cortese}
\affiliation[Southampton]
{School of Physics and Astronomy, University of Southampton, Southampton, SO17 1BJ, United Kingdom}
\author{Mattias Beck}
\affiliation[ETH]
{Institute of Quantum Electronics, ETH Z{\"u}rich, 8093 Z{\"u}rich, Switzerland}
\author{Simone De Liberato}
\affiliation[Southampton]
{School of Physics and Astronomy, University of Southampton, Southampton, SO17 1BJ, United Kingdom}
\author{J{\'e}r{\^o}me Faist}
\affiliation[ETH]
{Institute of Quantum Electronics, ETH Z{\"u}rich, 8093 Z{\"u}rich, Switzerland}
\author{Giacomo Scalari}
\affiliation[ETH]
{Institute of Quantum Electronics, ETH Z{\"u}rich, 8093 Z{\"u}rich, Switzerland}
\email{scalari@phys.ethz.ch}

\title[An \textsf{achemso} demo]
  {An engineered planar plasmonic reflector for polaritonic mode confinement}

\abbreviations{IR,NMR,UV}

\begin{document}

\begin{abstract}

It was recently demonstrated that, in deep subwavelength gap resonators coupled to two-dimensional electron gases, coupling to propagating plasmons can lead to energy leakage and prevent the formation of polaritonic resonances. This process, akin to Landau damping, limits the achievable field confinement and thus the value of light-matter coupling strength. In this work, we show how plasmonic subwavelength reflectors can be used to create an artificial energy stopband in the plasmon dispersion, confining them and enabling the recovery of the polaritonic resonances. Using this approach we demonstrate a normalized light-matter coupling ratio of $\frac{\Omega}{\omega} = 0.35 $ employing a single quantum well with a gap size of $\lambda/2400$ in vacuum.
\end{abstract}

\newpage
\section{Introduction}

The terahertz (THz) range is especially suited for the study of ultrastrong light-matter coupling phenomena \cite{forn2019ultrastrong,FriskKockum2019} since it combines material systems with large optical dipoles with extremely subwavelength resonant cavities obtained by exploiting metals in a lumped-circuit approach \cite{Jeannin2020,Geiser2012}. 
The Landau polariton platform\cite{Hagenmuller2010,Scalari2012} has proven especially successful in obtaining high values of the normalized coupling constant $\frac{\Omega}{\omega}>1$ \cite{Bayer2017}, extremely high cooperativity \cite{li2018vacuum, zhang2016collective}, as well as providing a testbed for investigating the influence of enhanced vacuum fields on the DC magnetotransport both in the linear \cite{paravicini2019magneto} and the integer Quantum Hall regimes \cite{Appugliese2022}. Several demonstrations of few-electron systems have been also realized in Landau polariton platforms \cite{Rajabali2022, keller2017few}.  

Reducing the modal volume of the electromagnetic resonator allows both to increase the strength of the light-matter coupling and to reduce the number of involved electrons \cite{Ballarini2019}, thus approaching the non-linear polaritonic regime. In standard nanophotonic platforms, both metallic \cite{Ciraci2012} and dielectric \cite{Gubbin2020}, this strategy is known to break down for resonator features small enough to excite propagative electronic excitation. In this case the standard local description of light-matter coupling fails and more accurate non-local approaches have to be used \cite{Fernandez-Dominguez2012,Gubbin2020B}.

A recent work \cite{Rajabali2021} demonstrated the  existence of a related effect in Landau polaritons due to the non-local excitation of propagative plasmons, limiting
the possibility of arbitrarily increasing electromagnetic confinement by reducing the cavity modal volume. In the aforementioned Landau-polariton paper, the magnetoplasmons which are collective inter-Landau level excitations are (ultra)strongly coupled to the  near-field of metamaterial complementary split-ring resonators (cSRR). The authors observed the progressive broadening and amplitude reduction of the upper polariton (UP) mode and partial disappearance of the lower polariton (LP) branch by reducing the gap of the coupled cSRRs below a critical length. According to our theoretical analysis, strongly subwavelength fields can excite a continuum of high-momenta propagative magnetoplasmons. These propagating modes act as loss channels and reduce the field confinement ultimately limiting the achievable field enhancement. As a result, certain polaritonic modes can broaden/disappear and the system enters a new regime of discrete-to-continuum strong coupling \cite{Cortese2019,Cortese2021}. In this work, we show how, by defining an engineered plasmonic mirror around the gap of the resonator we can again confine these propagating waves and retrieve well-defined polariton branches. 

\section{Results and Discussion}
\subsection{The reflector design}
\label{subsec:CST-refl-design}
\begin{figure}[h!]
\includegraphics[width=1\textwidth]{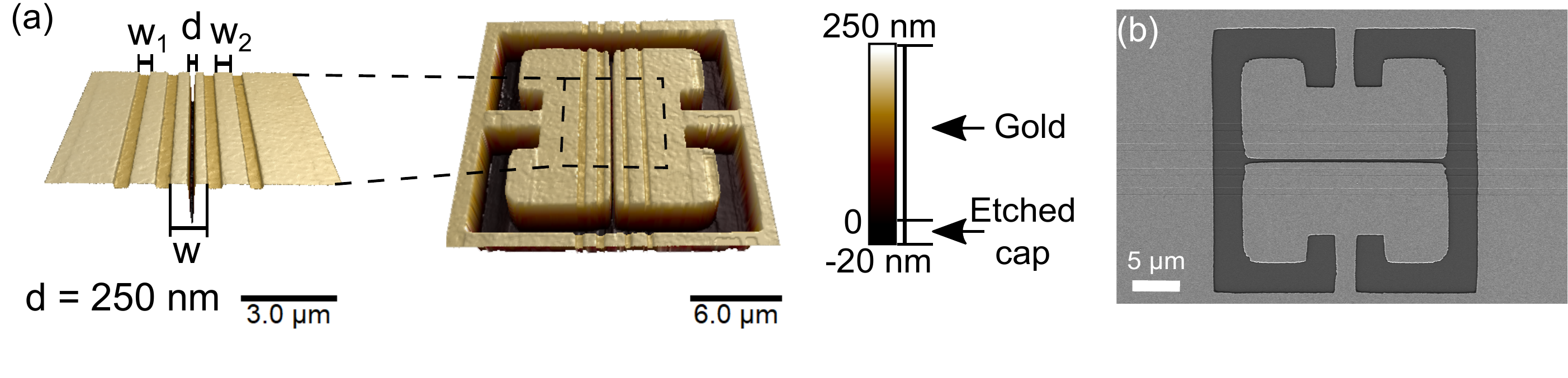}
\caption{\textbf{The scheme of the plasmonic reflector design for reconfinement of the broadened UP mode} (a) The 3D image, taken by atomic force microscopy, of the cSRR with a gap size of $d =\SI{250}{\nano\meter}$ on top a plasmonic reflector (the periodic trenches on the 2DEG). The parameters d, w$_1$, w$_2$, and w are $0.25$, $1$, $1$, and $\SI{1.5}{\micro\meter}$, respectively.(b) A scanning electron microscopy image of the same structure shown in panel (a). } 
\label{fig:schem-refl}
\end{figure}

In order to confine the broadened polaritonic mode whose energy is leaking out as a result of polaritonic non-local effects, we can introduce a plasmonic bandgap structure: a planar reflector in the magnetoplasmon propagation path, similar to plasmonic Bragg reflectors \cite{qu2016, song2016}. To build such a planar reflector for the magnetoplasmon waves, we need to introduce a modulation of the complex refractive index, i.e., of the dielectric function. One simple method is to introduce a modulation in the carrier density of the two-dimensional electron gas (2DEG) by structuring the $\SI{90}{\nano\meter}$-thick gallium arsenide (GaAs) cap layer on top of the quantum well channel, as illustrated in figure~\ref{fig:schem-refl}.

This structure, made by etching shallow trenches on both sides of the resonator gap, can define a stop-band in the transmission spectrum of the wave propagating across the 2DEG plane \cite{qu2016, song2016}. %By introducing a defect at the center of this periodic (2 periods) structure, a central mode can be also defined inside the stop-band. 
As a result, specifying such a periodic structure with a central defect below the sub-micron gap of the cSRR should allow retrieving the contrast in the amplitude and enhance the lifetime of the UP, effectively decoupling it from the lossy continuum of plasmonic modes. The three dimensional image of the fabricated resonator processed by atomic force microscopy is also exhibited in figure~\ref{fig:schem-refl}a indicating the important design parameters such as the resonator's gap (d), the central defect's width (w), and the width of the periodic structure of the reflector (w$_1$ and w$_2$).\\

\subsection{Landau polaritons in a plasmonic reflector}
\label{subsec:meas-cst-refl}
After a set of finite element simulation to find the right dimensions for reconfining the UP mode, a sample with a trench depth of $h = \SI{20}{\nano\meter}$, was fabricated (more details can be found in Methods section). Figure~\ref{fig:schem-refl}b shows a scanning electron microscopy image of the etched trenches in the 2DEG with a cSRR on top. The resonators were aligned and written on the etched substrate using electron-beam lithography.

\begin{figure}
\includegraphics[width=1\textwidth]{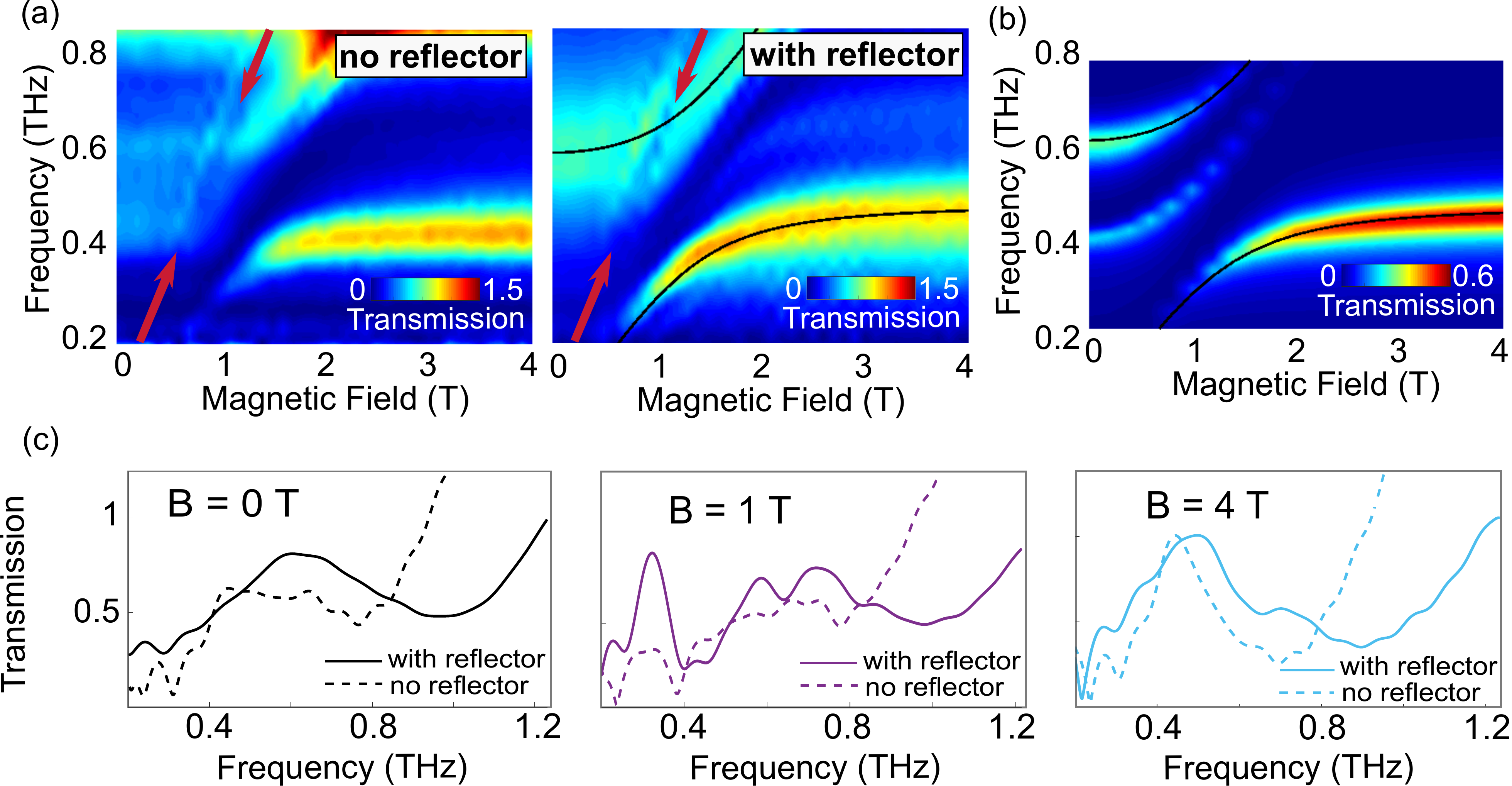}
\caption{\textbf{The cSRR with a gap size of 250 nm and the plasmonic reflector: experiment vs. simulation} (a) The measured transmission of the coupled cSRR to inter-Landau level transitions in a 2DEG without and with the plasmonic reflector (with h$~=\SI{20}{\nano\meter}$ etching depth). The one without the reflector illustrates the nonlocal effects (broadening of the modes) in its polaritonic modes while the one with the reflector has confined modes. The polariton branches from the measurement of the sample with reflector can be fitted by Hopfield model (black solid lines). The fitting indicates a normalized coupling of $\frac{\Omega_R}{\omega_0} = 35\%$. The red arrows mark the linear dispersions at multiple of the cyclotron frequency discussed in the main text. (b) Finite element simulation of the same resonator with the reflector in which the trenches are etched as deep as the entire cap layer thickness (h$~=\SI{90}{\nano\meter}$). A high-quality factor mode starting at $\SI{405}{\giga\hertz}$ at zero magnetic field appears only in the simulated colormap. The normalized coupling for the simulated spectra is $\frac{\Omega_R}{\omega_0} = 42\%$. (c) Sections of experimental data in panel (a) at three different values of the magnetic field, $B = 0$, $1$ (anti-crossing), and $\SI{4}{\tesla}$ with (solid) and without (dashed) the plasmonic reflector, confirm the retrieval of the broadened mode with the reflector.}
\label{fig:meas-reflect-250nm}
\end{figure}
\begin{figure}
\centering
\includegraphics[width=\textwidth]{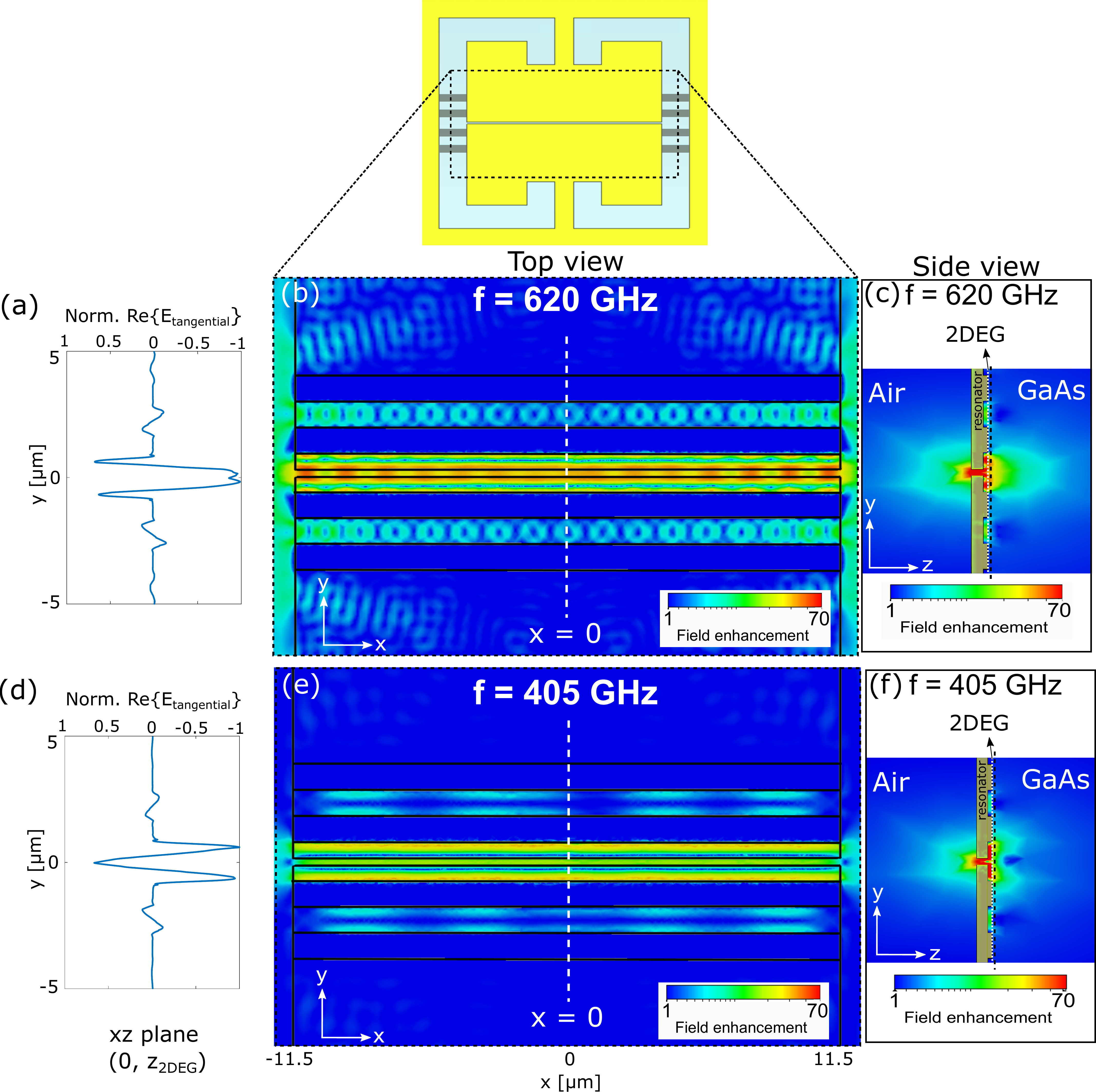}
\caption{\textbf{Field distribution of the high-quality factor mode and the UP mode} (a), (b), and (c): The tangential electric field distribution for the UP mode at $f = \SI{620}{\giga\hertz}$ at zero magnetic field. Panel (b) shows the x-y plane at $z = z_{2DEG}$. Panel (a) is a cross section of (b) (the real part) at $x = 0$ marked by dashed white line. Panel (c) illustrates the y-z plane at $x = 0$. The 2DEG is marked by dotted white line. The dashed black line indicates the place of the cross-section x-y plane for panel (b). (d), (e), and (f): similar to (a), (b), and (c) but for the high-quality factor mode at $f = \SI{405}{\giga\hertz}$ at zero magnetic field.}  
\label{fig:meas-reflect-250nm-400mode}
\end{figure}

Figure~\ref{fig:meas-reflect-250nm}a shows the transmission measurement for the cSRR with $\SI{250}{\nano\meter}$ gap size on a plasmonic reflector with $\SI{20}{\nano\meter}$-deep trenches. The measurements are conducted in a THz time-domain spectroscopy (TDS) setup at cryogenic temperature T = 2.7 K as a function of an external out-of-plane magnetic field swept between 0 and 4~T. In the THz-TDS setup, a pair of off-axis parabolic mirrors first are used to collimate and focus the incident THz beam from a photoconductive switch \cite{Madeo2010} on the sample. Then, through another pair of off-axis parabolic mirrors, the transmitted signal from the sample is collected, collimated, and focused on a zinc telluride (ZnTe) crystal. Ultimately, the THz signal is detected using an electro-optic detection scheme \cite{Gallot1999}. Detailed information about the THz-TDS setup is available in reference \cite{scalari2015}.\\
The transmission measurement without and with the reflector in figure~\ref{fig:meas-reflect-250nm}a shows the confinement of the UP and also the appearance of the LP mode at finite values of the magnetic field for the coupled cSRR on the reflector structure. There is no sign of magnetoplasmon dispersion in the measurement as the fitted LP (solid black line) converges to zero energy at zero magnetic field. For a better comparison between the polaritonic modes of the measurement without and with the reflector, section of the colormaps at three different values of the magnetic field, $B = 0$, $1$ (anti-crossing), and $\SI{4}{\tesla}$ are displayed in figure~\ref{fig:meas-reflect-250nm}c. 
\\Interestingly, the linear dispersions corresponding to optical transitions at multiples of the cyclotron frequency (indicated with red arrows in figure~\ref{fig:meas-reflect-250nm}a) are appearing in the transmission spectra of the sample with the reflector structure at the same place in the spectra of the sample without reflector which was also reported in our previous work~\cite{Rajabali2021}. These additional absorption lines are possible signatures of the breaking of the dipole approximation which results in the relaxation of the optical selection rules between the Landau levels. \\
We have also simulated the transmission spectrum of the coupled system with the reflector structure as a function of the magnetic field using the finite element method (figure~\ref{fig:meas-reflect-250nm}b). Given the difficulty of precisely modeling the modulation of the 2DEG's carrier due to the etching of the cap layer,
in the simulations the trenches are etched as deep as the entire cap layer thickness, $h = \SI{90}{\nano\meter}$.  The computed normalized coupling for the measured (figure~\ref{fig:meas-reflect-250nm}a) and the simulated (figure~\ref{fig:meas-reflect-250nm}b) spectrum of this coupled sample combined with the plasmonic reflector are $\frac{\Omega_R}{\omega_0} = 35\%$ and $\frac{\Omega_R}{\omega_0} = 42\%$, respectively. Besides the UP and LP modes, an additional high-quality factor mode starting from $\SI{405}{\giga\hertz}$ at zero magnetic field appears in the simulated transmission spectra, which is not visible in the experimental data. To assess the origin of this mode, the field distribution of the high-quality factor mode at $f = \SI{405}{\giga\hertz}$ and of the UP mode at $f = \SI{620}{\giga\hertz}$ are investigated at $\sim$zero magnetic field, shown in figure~\ref{fig:meas-reflect-250nm-400mode}. By comparing the tangential electric field in x-y plane in figure~\ref{fig:meas-reflect-250nm-400mode}a, \ref{fig:meas-reflect-250nm-400mode}b and \ref{fig:meas-reflect-250nm-400mode}d, \ref{fig:meas-reflect-250nm-400mode}e, the high-quality factor mode at $f = \SI{405}{\giga\hertz}$ seems to be a higher order mode. The full width half maximum (FWHM) of the central mode at $\SI{405}{\giga\hertz}$ and $\SI{620}{\giga\hertz}$ are $190$ and $\SI{660}{\nano\meter}$, respectively. The y-z plane cuts (figure~\ref{fig:meas-reflect-250nm-400mode}c and \ref{fig:meas-reflect-250nm-400mode}f) clearly displays that the high-quality factor mode does not couple to the resonator as there is no field confinement below the gap in the GaAs substrate below the 2DEG. On the contrary, the field distribution of the UP mode shows the field confinement below the resonator gap as a result of the coupling of the resonator to this mode. This high-quality factor mode which only appears in the simulation seems to have a wave vector comparable to the free electron wave vector. Hence, it easily dissipates and cannot appear in the measurement due to Landau damping~\cite{Khurgin2017}.\\
Another key point in the simulated electric field distribution is the electric field confinement in y-direction. Even though the presence of the reflector structure around the resonator gap can reconfine the UP mode, the electric field cannot be confined inside the gap of the resonator (d$~= \SI{250}{\nano\meter}$) and it is only confined to the central defect width (w$~= \SI{1.5}{\mu\meter}$). Thus, we can retrieve the ultrastrong coupling of our Landau polaritonic system but the highest achievable strength of the coupling will remain limited and the electromagnetic field confinement can be only reduced to the width of the central defect area. This explains why the achieved normalized coupling strength for the recovered polaritonic modes (figure~\ref{fig:meas-reflect-250nm}a) was a few percent lower than the expected normalized coupling strength for a coupled cSRR with a gap of d$~= \SI{250}{\nano\meter}$.

\newpage

\textbf{Conclusion}\par In this work, an engineered reflector design is proposed to retrieve the broadened polaritonic mode in systems with sub-micron size cavities coupled to inter-Landau level excitations in 2DEG. The initial broadening of the polaritonic mode originates from polaritonic nonlocality effects, discussed in our previous work~\cite{Rajabali2021}. The proposed design is based on periodic one-dimensional structures on both sides of the resonator gap to reflect the propagative magnetoplasmon waves and confine them in the gap of the resonator. %Although simulating the exact structure is not possible with finite element simulation but the simulation results can be useful for designing the geometrical parameters of the reflector. 
 After implementing such a structure around a coupled cSRR's gap with 250 nm width, in which polaritonic non-locality effect was observed, the polaritonic modes are retrieved. The recovered branches show a normalized coupling strength of $\Omega_R/\omega_0 = 35\%$ which is slightly lower than the expected coupling strength without considering the nonlocal effects. This is because the reconfined electromagnetic field is confined to the central width (w$~= \SI{1.5}{\mu\meter}$) of the proposed structure and not to the resonator's gap (d$~= \SI{250}{\nano\meter}$). The measurements are also verified by our finite element simulation.
Our result can be further improved by optimizing the reflector structure and can be also used for the direct resonators. This method can offer a solution to overcome the polaritonic nonlocal effects that introduce loss channels into the coupled systems and limit the enhancement of the coupling, however, the maximum achievable coupling strength saturates as the electromagnetic field confinement remains limited to the central width of the reflector structure.
\section{Methods}
\label{sec:methods}
To fabricate the plasmonic reflector, first, the trenches were defined using a direct laser writing lithography with Heidelberg DWL66+. The trenches were etched by a highly diluted etching solution ($slow$  $etchant$ $solution$: $H_2O$, $1:3$). The slow etchant solution itself was made of ($H_2SO_4:H_2O_2:H_2O$, $1:8:60$) with an etch rate of $\sim \SI{9}{\nano\meter/\second}$ for semi-insulating GaAs. The diluted solution had an etch rate of $\sim \SI{3}{\nano\meter/\second}$ for semi-insulating GaAs. Reducing the etch rate was done to increase the etching time and the accuracy of the etching to be able to etch only a few tens of nanometers. The etch rates were evaluated by a Dektak surface step profiler (with a $\SI{5}{\micro\meter}$ radius tip) and a scanning electron microscopy. After an aligning step, the cSRRs were fabricated on top of the etched trenches with electron beam lithography using a bilayer resist process with $\SI{450}{\nano\meter}$ 495K-PMMA-A4 and $\SI{90}{\nano\meter}$ Dow Corning XR-1541-006 electron beam negative resist. The lithography step was followed by deposition of $\SI{4}{\nano\meter}$ titanium and $\SI{200}{\nano\meter}$ of gold and a lift-off process.

\begin{acknowledgement}
S.R., J.E., M.B., J.F., and G.S. acknowledge financial support from ERC Grant No. 340975-MUSiC and the Swiss National Science Foundation (SNF) through the National Centre of Competence in Research Quantum Science and Technology (NCCR QSIT).\\
S.D.L. acknowledges the support of a University Research Fellowship of the Royal Society and of the Phillip Leverhulme prize of the Leverhulme Trust. S.D.L. and E.C. acknowledge funding from the RPG-2022-037 grant of the Leverhulme Trust. 

\end{acknowledgement}
%\bibliography{references}

\begin{thebibliography}{10}
\expandafter\ifx\csname url\endcsname\relax
  \def\url#1{\texttt{#1}}\fi
\expandafter\ifx\csname urlprefix\endcsname\relax\def\urlprefix{URL }\fi
\providecommand{\bibinfo}[2]{#2}
\providecommand{\eprint}[2][]{\url{#2}}

\bibitem[Forn-D{\'{\i}}az \latin{et~al.}(2019)Forn-D{\'{\i}}az, Lamata, Rico,
  Kono, and Solano]{forn2019ultrastrong}
Forn-D{\'{\i}}az,~P.; Lamata,~L.; Rico,~E.; Kono,~J.; Solano,~E. Ultrastrong
  coupling regimes of light-matter interaction. \emph{Reviews of Modern
  Physics} \textbf{2019}, \emph{91}\relax
\mciteBstWouldAddEndPuncttrue
\mciteSetBstMidEndSepPunct{\mcitedefaultmidpunct}
{\mcitedefaultendpunct}{\mcitedefaultseppunct}\relax
\EndOfBibitem
\bibitem[Frisk~Kockum \latin{et~al.}(2019)Frisk~Kockum, Miranowicz,
  De~Liberato, Savasta, and Nori]{FriskKockum2019}
Frisk~Kockum,~A.; Miranowicz,~A.; De~Liberato,~S.; Savasta,~S.; Nori,~F.
  {Ultrastrong coupling between light and matter}. \emph{{Nature Rev. Phys.}}
  \textbf{2019}, \emph{1}, 19--40\relax
\mciteBstWouldAddEndPuncttrue
\mciteSetBstMidEndSepPunct{\mcitedefaultmidpunct}
{\mcitedefaultendpunct}{\mcitedefaultseppunct}\relax
\EndOfBibitem
\bibitem[Jeannin \latin{et~al.}({2020})Jeannin, Bonazzi, Gacemi, Vasanelli, Li,
  Davies, Linfield, Sirtori, and Todorov]{Jeannin2020}
Jeannin,~M.; Bonazzi,~T.; Gacemi,~D.; Vasanelli,~A.; Li,~L.; Davies,~A.~G.;
  Linfield,~E.; Sirtori,~C.; Todorov,~Y. {Absorption Engineering in an
  Ultrasubwavelength Quantum System}. \emph{{Nano Lett.}} \textbf{{2020}},
  \emph{{20}}, {4430--4436}\relax
\mciteBstWouldAddEndPuncttrue
\mciteSetBstMidEndSepPunct{\mcitedefaultmidpunct}
{\mcitedefaultendpunct}{\mcitedefaultseppunct}\relax
\EndOfBibitem
\bibitem[Geiser \latin{et~al.}(2012)Geiser, Castellano, Scalari, Beck, Nevou,
  and Faist]{Geiser2012}
Geiser,~M.; Castellano,~F.; Scalari,~G.; Beck,~M.; Nevou,~L.; Faist,~J.
  {Ultrastrong Coupling Regime and Plasmon Polaritons in Parabolic
  Semiconductor Quantum Wells}. \emph{{Phys. Rev. Lett.}} \textbf{2012},
  \emph{108}, 106402\relax
\mciteBstWouldAddEndPuncttrue
\mciteSetBstMidEndSepPunct{\mcitedefaultmidpunct}
{\mcitedefaultendpunct}{\mcitedefaultseppunct}\relax
\EndOfBibitem
\bibitem[Hagenm\"uller \latin{et~al.}(2010)Hagenm\"uller, De~Liberato, and
  Ciuti]{Hagenmuller2010}
Hagenm\"uller,~D.; De~Liberato,~S.; Ciuti,~C. Ultrastrong coupling between a
  cavity resonator and the cyclotron transition of a two-dimensional electron
  gas in the case of an integer filling factor. \emph{Phys. Rev. B}
  \textbf{2010}, \emph{81}, 235303\relax
\mciteBstWouldAddEndPuncttrue
\mciteSetBstMidEndSepPunct{\mcitedefaultmidpunct}
{\mcitedefaultendpunct}{\mcitedefaultseppunct}\relax
\EndOfBibitem
\bibitem[Scalari \latin{et~al.}(2012)Scalari, Maissen, Tur{\v c}inkov{\'a},
  Hagenm{\"u}ller, De~Liberato, Ciuti, Reichl, Schuh, Wegscheider, Beck, and
  Faist]{Scalari2012}
Scalari,~G.; Maissen,~C.; Tur{\v c}inkov{\'a},~D.; Hagenm{\"u}ller,~D.;
  De~Liberato,~S.; Ciuti,~C.; Reichl,~C.; Schuh,~D.; Wegscheider,~W.; Beck,~M.;
  Faist,~J. {Ultrastrong Coupling of the Cyclotron Transition of a 2D Electron
  Gas to a THz Metamaterial}. \emph{{Science}} \textbf{2012}, \emph{335},
  1323--1326\relax
\mciteBstWouldAddEndPuncttrue
\mciteSetBstMidEndSepPunct{\mcitedefaultmidpunct}
{\mcitedefaultendpunct}{\mcitedefaultseppunct}\relax
\EndOfBibitem
\bibitem[Bayer \latin{et~al.}(2017)Bayer, Pozimski, Schambeck, Schuh, Huber,
  Bougeard, and Lange]{Bayer2017}
Bayer,~A.; Pozimski,~M.; Schambeck,~S.; Schuh,~D.; Huber,~R.; Bougeard,~D.;
  Lange,~C. {Terahertz Light-Matter Interaction beyond Unity Coupling
  Strength}. \emph{{Nano Lett.}} \textbf{2017}, \emph{17}, 6340--6344\relax
\mciteBstWouldAddEndPuncttrue
\mciteSetBstMidEndSepPunct{\mcitedefaultmidpunct}
{\mcitedefaultendpunct}{\mcitedefaultseppunct}\relax
\EndOfBibitem
\bibitem[Li \latin{et~al.}(2018)Li, Bamba, Zhang, Fallahi, Gardner, Gao, Lou,
  Yoshioka, Manfra, and Kono]{li2018vacuum}
Li,~X.; Bamba,~M.; Zhang,~Q.; Fallahi,~S.; Gardner,~G.~C.; Gao,~W.; Lou,~M.;
  Yoshioka,~K.; Manfra,~M.~J.; Kono,~J. Vacuum Bloch--Siegert shift in Landau
  polaritons with ultra-high cooperativity. \emph{Nature Photonics}
  \textbf{2018}, \emph{12}, 324--329\relax
\mciteBstWouldAddEndPuncttrue
\mciteSetBstMidEndSepPunct{\mcitedefaultmidpunct}
{\mcitedefaultendpunct}{\mcitedefaultseppunct}\relax
\EndOfBibitem
\bibitem[Zhang \latin{et~al.}(2016)Zhang, Lou, Li, Reno, Pan, Watson, Manfra,
  and Kono]{zhang2016collective}
Zhang,~Q.; Lou,~M.; Li,~X.; Reno,~J.~L.; Pan,~W.; Watson,~J.~D.; Manfra,~M.~J.;
  Kono,~J. Collective non-perturbative coupling of 2D electrons with
  high-quality-factor terahertz cavity photons. \emph{Nature Physics}
  \textbf{2016}, \emph{12}, 1005--1011\relax
\mciteBstWouldAddEndPuncttrue
\mciteSetBstMidEndSepPunct{\mcitedefaultmidpunct}
{\mcitedefaultendpunct}{\mcitedefaultseppunct}\relax
\EndOfBibitem
\bibitem[Paravicini-Bagliani \latin{et~al.}(2018)Paravicini-Bagliani,
  Appugliese, Richter, Valmorra, Keller, Beck, Bartolo, R\"{o}ssler, Ihn,
  Ensslin, Ciuti, Scalari, and Faist]{paravicini2019magneto}
Paravicini-Bagliani,~G.~L.; Appugliese,~F.; Richter,~E.; Valmorra,~F.;
  Keller,~J.; Beck,~M.; Bartolo,~N.; R\"{o}ssler,~C.; Ihn,~T.; Ensslin,~K.;
  Ciuti,~C.; Scalari,~G.; Faist,~J. Magneto-transport controlled by {Landau}
  polariton states. \emph{Nature Physics} \textbf{2018}, \emph{15},
  186--190\relax
\mciteBstWouldAddEndPuncttrue
\mciteSetBstMidEndSepPunct{\mcitedefaultmidpunct}
{\mcitedefaultendpunct}{\mcitedefaultseppunct}\relax
\EndOfBibitem
\bibitem[Appugliese \latin{et~al.}(2022)Appugliese, Enkner,
  Paravicini-Bagliani, Beck, Reichl, Wegscheider, Scalari, Ciuti, and
  Faist]{Appugliese2022}
Appugliese,~F.; Enkner,~J.; Paravicini-Bagliani,~G.~L.; Beck,~M.; Reichl,~C.;
  Wegscheider,~W.; Scalari,~G.; Ciuti,~C.; Faist,~J. Breakdown of topological
  protection by cavity vacuum fields in the integer quantum Hall effect.
  \emph{Science} \textbf{2022}, \emph{375}, 1030--1034\relax
\mciteBstWouldAddEndPuncttrue
\mciteSetBstMidEndSepPunct{\mcitedefaultmidpunct}
{\mcitedefaultendpunct}{\mcitedefaultseppunct}\relax
\EndOfBibitem
\bibitem[Rajabali \latin{et~al.}(2022)Rajabali, Markmann, J{\"o}chl, Beck,
  Lehner, Wegscheider, Faist, and Scalari]{Rajabali2022}
Rajabali,~S.; Markmann,~S.; J{\"o}chl,~E.; Beck,~M.; Lehner,~C.~A.;
  Wegscheider,~W.; Faist,~J.; Scalari,~G. An ultrastrongly coupled single
  terahertz meta-atom. \emph{Nature Communications} \textbf{2022}, \emph{13},
  2528\relax
\mciteBstWouldAddEndPuncttrue
\mciteSetBstMidEndSepPunct{\mcitedefaultmidpunct}
{\mcitedefaultendpunct}{\mcitedefaultseppunct}\relax
\EndOfBibitem
\bibitem[Keller \latin{et~al.}(2017)Keller, Scalari, Cibella, Maissen,
  Appugliese, Giovine, Leoni, Beck, and Faist]{keller2017few}
Keller,~J.; Scalari,~G.; Cibella,~S.; Maissen,~C.; Appugliese,~F.; Giovine,~E.;
  Leoni,~R.; Beck,~M.; Faist,~J. Few-Electron ultrastrong light-matter coupling
  at 300 GHz with nanogap hybrid LC microcavities. \emph{Nano letters}
  \textbf{2017}, \emph{17}, 7410--7415\relax
\mciteBstWouldAddEndPuncttrue
\mciteSetBstMidEndSepPunct{\mcitedefaultmidpunct}
{\mcitedefaultendpunct}{\mcitedefaultseppunct}\relax
\EndOfBibitem
\bibitem[Ballarini and {De Liberato}(2019)Ballarini, and {De
  Liberato}]{Ballarini2019}
Ballarini,~D.; {De Liberato},~S. {Polaritonics: from microcavities to
  sub-wavelength confinement}. \emph{{Nanophotonics}} \textbf{2019}, \emph{8},
  641--654\relax
\mciteBstWouldAddEndPuncttrue
\mciteSetBstMidEndSepPunct{\mcitedefaultmidpunct}
{\mcitedefaultendpunct}{\mcitedefaultseppunct}\relax
\EndOfBibitem
\bibitem[Cirac{\`\i} \latin{et~al.}(2012)Cirac{\`\i}, Hill, Mock, Urzhumov,
  Fern{\'a}ndez-Dom{\'\i}nguez, Maier, Pendry, Chilkoti, and Smith]{Ciraci2012}
Cirac{\`\i},~C.; Hill,~R.~T.; Mock,~J.~J.; Urzhumov,~Y.;
  Fern{\'a}ndez-Dom{\'\i}nguez,~A.~I.; Maier,~S.~A.; Pendry,~J.~B.;
  Chilkoti,~A.; Smith,~D.~R. {Probing the Ultimate Limits of Plasmonic
  Enhancement}. \emph{{Science}} \textbf{2012}, \emph{337}, 1072--1074\relax
\mciteBstWouldAddEndPuncttrue
\mciteSetBstMidEndSepPunct{\mcitedefaultmidpunct}
{\mcitedefaultendpunct}{\mcitedefaultseppunct}\relax
\EndOfBibitem
\bibitem[Gubbin and De~Liberato(2020)Gubbin, and De~Liberato]{Gubbin2020}
Gubbin,~C.~R.; De~Liberato,~S. {Optical Nonlocality in Polar Dielectrics}.
  \emph{{Phys. Rev. X}} \textbf{2020}, \emph{10}, 021027\relax
\mciteBstWouldAddEndPuncttrue
\mciteSetBstMidEndSepPunct{\mcitedefaultmidpunct}
{\mcitedefaultendpunct}{\mcitedefaultseppunct}\relax
\EndOfBibitem
\bibitem[Fern\'andez-Dom\'{\i}nguez
  \latin{et~al.}(2012)Fern\'andez-Dom\'{\i}nguez, Wiener, Garc\'{\i}a-Vidal,
  Maier, and Pendry]{Fernandez-Dominguez2012}
Fern\'andez-Dom\'{\i}nguez,~A.~I.; Wiener,~A.; Garc\'{\i}a-Vidal,~F.~J.;
  Maier,~S.~A.; Pendry,~J.~B. Transformation-Optics Description of Nonlocal
  Effects in Plasmonic Nanostructures. \emph{Phys. Rev. Lett.} \textbf{2012},
  \emph{108}, 106802\relax
\mciteBstWouldAddEndPuncttrue
\mciteSetBstMidEndSepPunct{\mcitedefaultmidpunct}
{\mcitedefaultendpunct}{\mcitedefaultseppunct}\relax
\EndOfBibitem
\bibitem[Gubbin and De~Liberato(2020)Gubbin, and De~Liberato]{Gubbin2020B}
Gubbin,~C.~R.; De~Liberato,~S. Impact of phonon nonlocality on nanogap and
  nanolayer polar resonators. \emph{Phys. Rev. B} \textbf{2020}, \emph{102},
  201302\relax
\mciteBstWouldAddEndPuncttrue
\mciteSetBstMidEndSepPunct{\mcitedefaultmidpunct}
{\mcitedefaultendpunct}{\mcitedefaultseppunct}\relax
\EndOfBibitem
\bibitem[Rajabali \latin{et~al.}(2021)Rajabali, Cortese, Beck, De~Liberato,
  Faist, and Scalari]{Rajabali2021}
Rajabali,~S.; Cortese,~E.; Beck,~M.; De~Liberato,~S.; Faist,~J.; Scalari,~G.
  {Polaritonic nonlocality in light--matter interaction}. \emph{{Nature Phot.}}
  \textbf{2021}, \emph{15}, 690--695\relax
\mciteBstWouldAddEndPuncttrue
\mciteSetBstMidEndSepPunct{\mcitedefaultmidpunct}
{\mcitedefaultendpunct}{\mcitedefaultseppunct}\relax
\EndOfBibitem
\bibitem[Cortese \latin{et~al.}(2019)Cortese, Carusotto, Colombelli, and
  De~Liberato]{Cortese2019}
Cortese,~E.; Carusotto,~I.; Colombelli,~R.; De~Liberato,~S. {Strong coupling of
  ionizing transitions}. \emph{{Optica}} \textbf{2019}, \emph{6},
  354--361\relax
\mciteBstWouldAddEndPuncttrue
\mciteSetBstMidEndSepPunct{\mcitedefaultmidpunct}
{\mcitedefaultendpunct}{\mcitedefaultseppunct}\relax
\EndOfBibitem
\bibitem[Cortese and De~Liberato(2022)Cortese, and De~Liberato]{Cortese2021}
Cortese,~E.; De~Liberato,~S. Exact solution of polaritonic systems with
  arbitrary light and matter frequency-dependent losses. \emph{The Journal of
  Chemical Physics} \textbf{2022}, \emph{156}, 084106\relax
\mciteBstWouldAddEndPuncttrue
\mciteSetBstMidEndSepPunct{\mcitedefaultmidpunct}
{\mcitedefaultendpunct}{\mcitedefaultseppunct}\relax
\EndOfBibitem
\bibitem[Qu \latin{et~al.}(2016)Qu, Song, Xia, Liang, Tang, Hu, and
  Wang]{qu2016}
Qu,~S.; Song,~C.; Xia,~X.; Liang,~X.; Tang,~B.; Hu,~Z.-D.; Wang,~J. {Detuned
  plasmonic Bragg grating sensor based on a defect metal-insulator-metal
  waveguide}. \emph{{Sensors}} \textbf{2016}, \emph{16}, 784\relax
\mciteBstWouldAddEndPuncttrue
\mciteSetBstMidEndSepPunct{\mcitedefaultmidpunct}
{\mcitedefaultendpunct}{\mcitedefaultseppunct}\relax
\EndOfBibitem
\bibitem[Song \latin{et~al.}(2016)Song, Xia, Hu, Liang, and Wang]{song2016}
Song,~C.; Xia,~X.; Hu,~Z.-D.; Liang,~Y.; Wang,~J. {Characteristics of plasmonic
  Bragg reflectors with graphene-based silicon grating}. \emph{{Nanoscale
  research lett.}} \textbf{2016}, \emph{11}, 1--8\relax
\mciteBstWouldAddEndPuncttrue
\mciteSetBstMidEndSepPunct{\mcitedefaultmidpunct}
{\mcitedefaultendpunct}{\mcitedefaultseppunct}\relax
\EndOfBibitem
\bibitem[Madeo \latin{et~al.}({2010})Madeo, Jukam, Oustinov, Rosticher,
  Rungsawang, Tignon, and Dhillon]{Madeo2010}
Madeo,~J.; Jukam,~N.; Oustinov,~D.; Rosticher,~M.; Rungsawang,~R.; Tignon,~J.;
  Dhillon,~S.~S. {Frequency tunable terahertz interdigitated photoconductive
  antennas}. \emph{{Elec. Lett.}} \textbf{{2010}}, \emph{{46}},
  {611--U25}\relax
\mciteBstWouldAddEndPuncttrue
\mciteSetBstMidEndSepPunct{\mcitedefaultmidpunct}
{\mcitedefaultendpunct}{\mcitedefaultseppunct}\relax
\EndOfBibitem
\bibitem[Gallot and Grischkowsky(1999)Gallot, and Grischkowsky]{Gallot1999}
Gallot,~G.; Grischkowsky,~D. {Electro-optic detection of terahertz radiation}.
  \emph{{J. Opt. Soc. Am. B}} \textbf{1999}, \emph{16}, 1204--1212\relax
\mciteBstWouldAddEndPuncttrue
\mciteSetBstMidEndSepPunct{\mcitedefaultmidpunct}
{\mcitedefaultendpunct}{\mcitedefaultseppunct}\relax
\EndOfBibitem
\bibitem[Scalari \latin{et~al.}(2015)Scalari, Maissen, Cibella, Leoni, Reichl,
  Wegscheider, Beck, and Faist]{scalari2015}
Scalari,~G.; Maissen,~C.; Cibella,~S.; Leoni,~R.; Reichl,~C.; Wegscheider,~W.;
  Beck,~M.; Faist,~J. {THz ultrastrong light-matter coupling}. \emph{{II Nuovo
  Saggiatore}} \textbf{2015}, \emph{31, 3-4}, 4--14\relax
\mciteBstWouldAddEndPuncttrue
\mciteSetBstMidEndSepPunct{\mcitedefaultmidpunct}
{\mcitedefaultendpunct}{\mcitedefaultseppunct}\relax
\EndOfBibitem
\bibitem[Khurgin \latin{et~al.}(2017)Khurgin, Tsai, Tsai, and Sun]{Khurgin2017}
Khurgin,~J.; Tsai,~W.-Y.; Tsai,~D.~P.; Sun,~G. {Landau Damping and Limit to
  Field Confinement and Enhancement in Plasmonic Dimers}. \emph{{ACS
  Photonics}} \textbf{2017}, \emph{4}, 2871--2880\relax
\mciteBstWouldAddEndPuncttrue
\mciteSetBstMidEndSepPunct{\mcitedefaultmidpunct}
{\mcitedefaultendpunct}{\mcitedefaultseppunct}\relax
\EndOfBibitem
\end{thebibliography}

\end{document}